\documentclass[12pt,a4paper]{article}

\usepackage[utf8]{inputenc}
\usepackage[T1]{fontenc}
\usepackage{times}
\usepackage[margin=1in]{geometry}
\usepackage{setspace}
\usepackage{graphicx}
\usepackage{booktabs}
\usepackage{multirow}
\usepackage{array}
\usepackage{longtable}
\usepackage{amsmath}
\usepackage{amssymb}
\usepackage[hyphens]{url}
\usepackage{listings}
\usepackage{xcolor}
\usepackage{float}
\newfloat{listing}{htbp}{lol}
\floatname{listing}{Listing}
\usepackage{etoolbox}
\usepackage[ruled,vlined,linesnumbered]{algorithm2e}
\SetKwInput{KwInput}{Input}
\SetKwInput{KwOutput}{Output}
\let\origtextsc\textsc
\newcommand{\algotextsc}[1]{\textnormal{\origtextsc{#1}}}
\AtBeginEnvironment{algorithm}{\let\textsc\algotextsc}
\usepackage[style=apa,backend=biber,natbib=true]{biblatex}
\usepackage[colorlinks=true,linkcolor=black,citecolor=black,urlcolor=black]{hyperref}

\title{Time travel for knowledge graphs: live queries over RDF change histories}

\author{
    Arcangelo Massari \and Silvio Peroni
}

\date{}

\begin{document}

\maketitle

\noindent Department of Classical Philology and Italian Studies, University of Bologna, Via Zamboni 32, 40126 Bologna, Italy

\vspace{1em}

\noindent\textbf{Corresponding author:} Arcangelo Massari\\
Email: arcangelo.massari@unibo.it

\vspace{0.5em}

\noindent\textbf{ORCID:} Arcangelo Massari: 0000-0002-8420-0696; Silvio Peroni: 0000-0003-0530-4305

\newpage

\begin{abstract}
Performing time-traversal queries on RDF datasets remains unsupported in the most extensive knowledge graphs. Existing solutions either require offline ingestion, which prevents concurrent querying and updating, or operate live but with limited query coverage or triplestore dependency. This article presents the Time Agnostic Library, a Python library for performing temporal SPARQL queries live on any SPARQL-compliant triplestore, supporting all six temporal retrieval needs identified in the literature and concurrent updates. The methodology builds on the OpenCitations Data Model (OCDM), which records provenance using the Provenance Ontology (PROV-O) and SPARQL UPDATE operations. The library supports version materialization, single-version and cross-version structured queries, delta materialization, and single-delta and cross-delta structured queries over multi-triple patterns. Evaluation on the BEAR-B benchmark shows sub-linear scaling in both execution time and memory consumption as the number of versions increases. While preprocessing-based systems such as OSTRICH achieve faster query times, they require offline ingestion and cannot handle concurrent data updates. Against R43ples, the closest live system in architecture, the Time Agnostic Library is faster across all query types.

\vspace{1em}
\end{abstract}

\newpage

\section{Introduction}

Assessing data reliability requires provenance information: who produced the data, when, and from which primary source. Moreover, data changes over time, either due to the natural evolution of concepts or to error correction. The latest version of a dataset may not be the most accurate. These phenomena affect the Web of Data, as shown by the Dynamic Linked Data Observatory, which reported modifications to about 38\% of the 86,696 RDF documents monitored for 29 weeks and the permanent disappearance of 5\% \citep{kaferObservingLinkedData2013}.

Notwithstanding these premises, performing SPARQL time-traversal queries on previous states of RDF entities together with provenance information remains unsupported in the most extensive RDF datasets. DBpedia relies on independent dump-based versioning rather than recording provenance or changes in RDF \citep{orlandiModellingProvenanceDBpedia2011,umbrichDatasetDynamicsChange2010}. Wikidata supports SPARQL queries over entities temporally annotated via its Wikibase-specific RDF data model, while its edit history is stored outside RDF \citep{dooleyLinkedDataWikidata2019}. A dedicated History Query Service was developed to convert Wikidata's JSON revision dumps into RDF named graphs and index them for SPARQL access, but has since been shut down \citep{pellissiertanonQueryingEditHistory2019}. YAGO 4.5 uses RDF-star annotations to record the real-world temporal validity of facts (for example, the period during which a political role was held), which serves a different purpose from tracking editorial changes to the dataset itself \citep{suchanekYAGO45Large2024}.

The main reason for this heterogeneity is that the founding technologies of the Semantic Web, namely SPARQL and RDF, did not initially provide an effective mechanism for annotating statements with metadata. This gap led to the introduction of numerous metadata representation models, none of which became widely accepted standards for tracking both provenance and changes to RDF entities \citep{massariRepresentingProvenanceTrack2025}. The RDF 1.2 specification, currently a W3C Working Draft, introduces native support for statement-level annotations through triple terms and reifiers \citep{kelloggRDF12Concepts2026}, addressing this syntactic limitation. However, a standard annotation mechanism does not, by itself, provide a model for tracking changes and provenance, nor does it provide software for performing temporal queries on historical data.

In the past, some software was developed to perform temporal queries on RDF datasets, enabling the reconstruction of the status of a particular entity at a given time. However, as far as we know, all existing solutions require an offline ingestion phase to preprocess and index RDF data into dedicated storage structures before queries can be executed \citep{cerdeira-penaSelfIndexingRDFArchives2016,imVersionManagementFramework2012,neumannXRDF3XFastQuerying2010,pellissiertanonQueryingEditHistory2019,taelmanOSTRICHVersionedRandomAccess2018,pelgrinGLENDAQueryingRDF2023,pelgrinExpressiveQueryingScalable2025}. During this ingestion phase, the store cannot be queried, and once built, data cannot be updated without repeating the ingestion process. This requirement is impractical for linked open datasets that constantly receive many updates, such as Wikidata.

More broadly, preprocessing is impractical for any RDF data editor that integrates provenance and change tracking, since such editors require live access to the history of the underlying data. HERITRACE \citep{massariHERITRACEUserFriendlySemantic2025} is one example of an editor with this requirement. Conversely, software operating on the fly either does not support all query types \citep{noyPromptdiffFixedPointAlgorithm2002}, or supports them non-generically by imposing a custom database \citep{graubeOpenSemanticRevision2016} or a specific triplestore \citep{arndtDecentralizedCollaborativeKnowledge2019,sandeRWbaseGitTriples2013}.

This article introduces a methodology and a Python library that enable all time-related retrieval functionalities identified by \citet{fernandezEvaluatingQueryStorage2016} in a live setting, supporting SPARQL queries with full basic graph patterns. Moreover, data can be stored on any RDF-compliant storage system (for example, RDF-serialized textual files and triplestores) when the provenance and data changes are tracked according to the OpenCitations Data Model \citep{daquinoOpenCitationsDataModel2020}.

The rest of the article is organized as follows. \hyperref[sec:temporal-queries]{Section 2} reviews the literature on temporal query typologies. \hyperref[sec:storage-paradigms]{Section 3} examines storage paradigms for dynamic linked open data. \hyperref[sec:ocdm-approach]{Section 4} describes provenance and change tracking in the OpenCitations Data Model. \hyperref[sec:time-agnostic-library]{Section 5} presents the Time Agnostic Library implementation. \hyperref[sec:evaluation]{Section 6} reports the benchmark results on execution times and memory consumption.

\section{Temporal query typologies for RDF datasets}
\label{sec:temporal-queries}

\citet{fernandezEvaluatingQueryStorage2016} established a classification for temporal queries on RDF archives at two levels. The first level identifies six retrieval needs, categorized by focus (version or delta) and type (materialization or structured query). The second level defines five query atoms, formal operations that take a SPARQL query $Q$ as a parameter. The two levels reuse the same terms with different meanings: for instance, ``version materialization'' at the retrieval needs level denotes the reconstruction of a complete dataset snapshot, whereas the corresponding query atom $Mat(Q, V_i) = [[Q]]_{V_i}$ denotes the evaluation of a query on a specific version, which corresponds to a single-version structured query at the retrieval needs level. To avoid ambiguity, this article uses the terminology of retrieval needs throughout. In the following, $V_{i}$ denotes the state of the dataset at version $i$, and $[[Q]]_{V_{i}}$ denotes the bag of solution mappings produced by evaluating SPARQL query $Q$ on $V_{i}$.

\textbf{Version materialization (VM)} retrieves the complete state of the dataset at a specific version $i$, that is, $V_{i}$. For example, ``Retrieve the state of the dataset in 2014''.

\textbf{Single-version structured query (SV)} retrieves the results of a SPARQL query targeted at a specific version: $SV\left(Q, V_{i}\right)=[[Q]]_{V_{i}}$. For example, ``Which David Shotton's papers were featured in the dataset in 2014?''.

\textbf{Cross-version structured query (CV)}, also called time-traversal query, retrieves the results of a SPARQL query targeted at multiple versions: $CV\left(Q, V_{i}, V_{j}\right)=SV\left(Q, V_{i}\right) \bowtie SV\left(Q, V_{j}\right)$. For example, ``Which David Shotton's papers were featured in the dataset in 2013 and in 2014?''.

\textbf{Delta materialization (DM)} retrieves the differences between two versions $V_{i}$ and $V_{j}$: $DM\left(V_{i}, V_{j}\right)=(\Delta^+, \Delta^-)$, with $\Delta^+=V_{j} \setminus V_{i}$ and $\Delta^-=V_{i} \setminus V_{j}$. For example, ``What changed in the dataset between 2013 and 2014?''.

\textbf{Single-delta structured query (SD)} retrieves the change-sets of a SPARQL query's results between two versions: $SD\left(Q, V_{i}, V_{j}\right)=(\Delta^+, \Delta^-)$, with $\Delta^+=[[Q]]_{V_{j}} \setminus [[Q]]_{V_{i}}$ and $\Delta^-=[[Q]]_{V_{i}} \setminus [[Q]]_{V_{j}}$. For example, ``What are the additions and removals of David Shotton's papers between 2013 and 2014?''.

\textbf{Cross-delta structured query (CD)} retrieves the change-sets of a SPARQL query's results across multiple version pairs: $CD\left(Q, V_{i}, V_{j}, V_{m}\right)=SD\left(Q, V_{i}, V_{j}\right) \bowtie SD\left(Q, V_{j}, V_{m}\right)$. For example, ``Track the evolution of David Shotton's papers across multiple version transitions in the dataset''.

Extensions of SPARQL exist to support queries on time-aware RDF datasets, but they either extend SPARQL syntax with temporal constructs, such as $\tau$-SPARQL \citep{tappoletAppliedTemporalRDF2009}, T-SPARQL \citep{grandiTSPARQLTSQL2likeTemporal2010}, and AnQL \citep{zimmermannGeneralFrameworkRepresenting2012}, or only work on purpose-built databases, such as SPARQL$^{T}$ on the RDF-TX system \citep{zanioloUserfriendlyTemporalQueries2018}.

SPARQL-LTL \citep{fiondaGizeTimeWarp2016} extends SPARQL with an algorithm for rewriting queries into standard SPARQL, requiring triples annotated with revision numbers and available as named graphs. The GiZe demonstration prototype is the only known implementation, and no publicly maintained version is available.

The methodology introduced in this article differs from all the above approaches in how temporal logic is handled: rather than extending the query language with temporal constructs or requiring a purpose-built database, the Time Agnostic Library relies on temporal information in data mapped according to the OCDM data model and reconstructs states programmatically. Temporal SPARQL extensions require either native support in triplestores or an external processing layer that translates extended queries into standard SPARQL. Since no temporal SPARQL extension has been adopted by mainstream triplestores or standardized by the W3C, the former path remains unavailable, and the latter results in the same architectural pattern as the approach presented here, with the addition of a non-standard query syntax.

\section{Storage paradigms for dynamic linked open data}
\label{sec:storage-paradigms}

Various archiving policies have been developed to store and query dynamic RDF datasets, namely independent copies, change-based, and timestamp-based policies \citep{pelgrinFullyfledgedArchivingRDF2021}, as well as fragment-based policies \citep{arndtDecentralizedCollaborativeKnowledge2019}.

\textbf{Independent copies} consist of storing each version separately. This represents the most straightforward model to implement and allows performing version materialization, single-version queries, and cross-version queries easily. However, this approach requires a massive amount of storage space. Furthermore, given the different versions of the statements, additional diff mechanisms are required to identify what changed. Nevertheless, this remains the archiving policy adopted by most systems and knowledge bases, such as DBpedia \citep{lehmannDBpediaLargescaleMultilingual2015}, Wikidata \citep{vrandecicWikidataFreeCollaborative2014}, and YAGO \citep{suchanekYAGO45Large2024}.

Among the earliest version control systems for RDF was SemVersion \citep{volkelSemVersionRDFbased2006}, designed for RDF models and RDF-based ontology languages such as RDFS and OWL. It saves each version in a separate snapshot, and differences are calculated on the fly. SemVersion supports version materialization and delta materialization, but not via SPARQL, because SPARQL became a W3C Recommendation in 2008, and SemVersion has not been updated since 2006.

\textbf{Change-based policy} was introduced to address scalability issues arising from the independent copies approach. It consists of saving only the deltas between the two versions. For this reason, delta materialization is costless. The drawback is that additional computational costs for delta propagation are required to support version-focused queries.

An early proposal of this approach relied on a Relational Database Management System (RDBMS) to store the original dataset and the deltas between two consecutive versions \citep{imVersionManagementFramework2012}. To improve performance, deltas are preprocessed and duplicated, or unnecessary modifications are deleted. There is no support for SPARQL, and queries must be formulated in SQL.

A concrete implementation of a change-based policy is R\&Wbase \citep{sandeRWbaseGitTriples2013}, a version control system inspired by Git but designed for RDF. Additions and deletions are stored using distinct context identifiers in a quad-store, and SPARQL queries are supported on versioned data. However, this model is not fully semantic, since it requires hash tables to map revisions with change-sets.

R43ples \citep{graubeOpenSemanticRevision2016} is inspired by R\&WBase and replaces its non-semantic mapping components with the Revision Management Ontology (RMO), a fully semantic model that records change-sets and provenance metadata in separate named graphs using PROV-O. R43ples extends SPARQL with formalized keywords for revision management (\texttt{REVISION}, \texttt{BRANCH}, \texttt{TAG}, \texttt{MERGE}) and acts as a proxy in front of a triplestore, rewriting extended queries into standard SPARQL to reconstruct requested revisions from stored change-sets. Although the architecture defines a generic triplestore interface, the implementation is coupled to Apache Jena: the interface methods return Jena-specific types (\texttt{ResultSet}, \texttt{Model}), and the only functional backend is Jena TDB. Past states are reconstructed by walking the revision chain from HEAD backward to the requested revision, executing a SPARQL query against the underlying triplestore at each step. The open-source project has seen no development since 2019.

A more recent change-based approach uses RDF-star for triple-level provenance tracking \citep{dibowskiFullTraceabilityProvenance2024}. A provenance engine intercepts SPARQL UPDATE queries and transforms them into SPARQL-star INSERT DATA operations on a separate provenance knowledge graph. Changes and their provenance are represented using PROV-STAR, an extension of PROV-O that introduces three classes for triple change sets (generation, invalidation, and their abstract parent). Past versions can be restored via a single SPARQL-star CONSTRUCT query that selects all triples generated before a given timestamp and filters out those invalidated before it. The approach requires SPARQL-star support in the underlying triplestore. Unfortunately, as far as we know, the provenance engine implementation has not been released as open-source software, preventing independent replication and evaluation of the system.

\textbf{Timestamp-based policy} annotates each triple with its transaction time, that is, the timestamp of the version in which that statement was in the dataset.

x-RDF-3X \citep{neumannXRDF3XFastQuerying2010} is a database for RDF designed to manage high-frequency online updates, versioning, time-traversal queries, and transactions. The triples are never deleted but are annotated with two fields: the insertion and deletion timestamps, with the deletion timestamp having a zero value for currently living versions. Afterward, updates are saved in a separate workspace and merged into various indexes at occasional savepoints. x-RDF-3X supports version materialization, single-version queries, and cross-version queries. However, to the best of our knowledge, its source code is not publicly available.

v-RDFCSA \citep{cerdeira-penaSelfIndexingRDFArchives2016} and its generalization, v-RDF-SI \citep{cerdeira-penaCompressedQueryableSelfindexes2024}, use a similar strategy but excel at reducing space requirements by compressing both the RDF archive and the timestamps associated with triples. Both systems support only version materialization, delta materialization, and version queries on single triple patterns, without full SPARQL or dynamic ingestion of new versions. Similarly, their source code is not publicly available, preventing independent replication.

Dydra \citep{andersonRDFGraphStores2019} is a timestamp-based system that stores each quad together with the identifiers of the transactions that added and removed it. It operates as a live system: each mutation via SPARQL UPDATE or the Graph Store Protocol automatically creates a new revision, identified by a UUID. Dydra extends SPARQL with a \texttt{REVISION} clause analogous to \texttt{GRAPH}, allowing queries to target specific revisions, iterate over all revisions, or reference the previous revision. It supports all query types. Unlike R43ples, which stores its provenance metadata in RDF via the Revision Management Ontology, Dydra encodes versioning information within its proprietary B+ tree indexes; this metadata is not represented in RDF and is not accessible independently of the Dydra storage engine. The system is therefore not triplestore-agnostic. Additionally, its source code is not publicly available; access is only available through the Dydra commercial cloud service.

\textbf{Fragment-based approach} avoids reconstructing versions via deltas by saving only fragments of what changed. Different granularity levels are possible, depending on the requirements: a graph, a subgraph, or an entity.

Like R\&Wbase, Quit Store \citep{arndtDecentralizedCollaborativeKnowledge2019} builds on Git for RDF version control. Each named graph is serialized to an N-Triples file, and every SPARQL UPDATE automatically creates a Git commit that records the changes. Provenance metadata are generated in RDF using PROV-O, mapping commits to \texttt{prov:Activity} instances and dataset revisions to \texttt{prov:Entity} instances, accessible through a dedicated SPARQL endpoint. The system uses standard SPARQL 1.1 without proprietary extensions. However, Quit Store does not support temporal queries: the SPARQL endpoint exposes only the current version of the dataset, and there is no mechanism to query past states. The entire dataset must fit in memory, as it is held in an rdflib-based in-memory quad store.

\textbf{Hybrid storage policies} combine multiple strategies. OSTRICH \citep{taelmanOSTRICHVersionedRandomAccess2018} combines the independent copies, change-based, and timestamp-based strategies: it stores the initial version as an HDT snapshot and represents subsequent versions as aggregated deltas relative to that snapshot, while annotating each triple with the version numbers in which it exists. In the original design, only a single snapshot at version~0 is used, leading to delta sizes and ingestion times growing as versions accumulate. \citet{pelgrinExpressiveQueryingScalable2025} extended this architecture with configurable snapshot-creation strategies that materialize intermediate snapshots, each of which starts a new delta chain. All version data must be ingested offline as N-Triples changeset files before the store can be queried, and individual triples cannot be inserted or deleted after ingestion. OSTRICH resolves version materialization, delta materialization, and version queries for single triple patterns via direct index lookups, without reconstructing full dataset states. However, it does not accept arbitrary SPARQL queries: only individual triple patterns are supported. GLENDA \citep{pelgrinGLENDAQueryingRDF2023} addresses this limitation by placing the Comunica query engine \citep{taelman_iswc_resources_comunica_2018} on top of OSTRICH. Given a SPARQL query targeting a specific version, Comunica decomposes it into individual triple patterns, resolves each one against the OSTRICH store for that version, and joins the results. Like OSTRICH, GLENDA requires offline ingestion and does not support live updates.

TailR \citep{meinhardtTailRPlatformPreserving2015} adopts a hybrid, independent copies/change-based approach: it stores periodic snapshots with deltas, grouped by resource (subject). It supports only version materialization at the individual resource level, retrievable via the Memento protocol \citep{jonesInteroperabilityAccessingVersions2021} (HTTP content-negotiation via the Accept-Datetime header), without SPARQL support. Versioning metadata are stored in a relational database (MariaDB), not in RDF. The source code, originally published on GitHub, is no longer available.

Finally, the OpenCitations Data Model \citep{daquinoOpenCitationsDataModel2020} adopts a hybrid approach that combines change-based and timestamp-based methods to represent provenance and changes in RDF, as detailed in the following section.

\section{Provenance and change tracking in the OpenCitations Data Model}
\label{sec:ocdm-approach}

\begin{figure}[htbp]
\centering
\includegraphics[width=\textwidth]{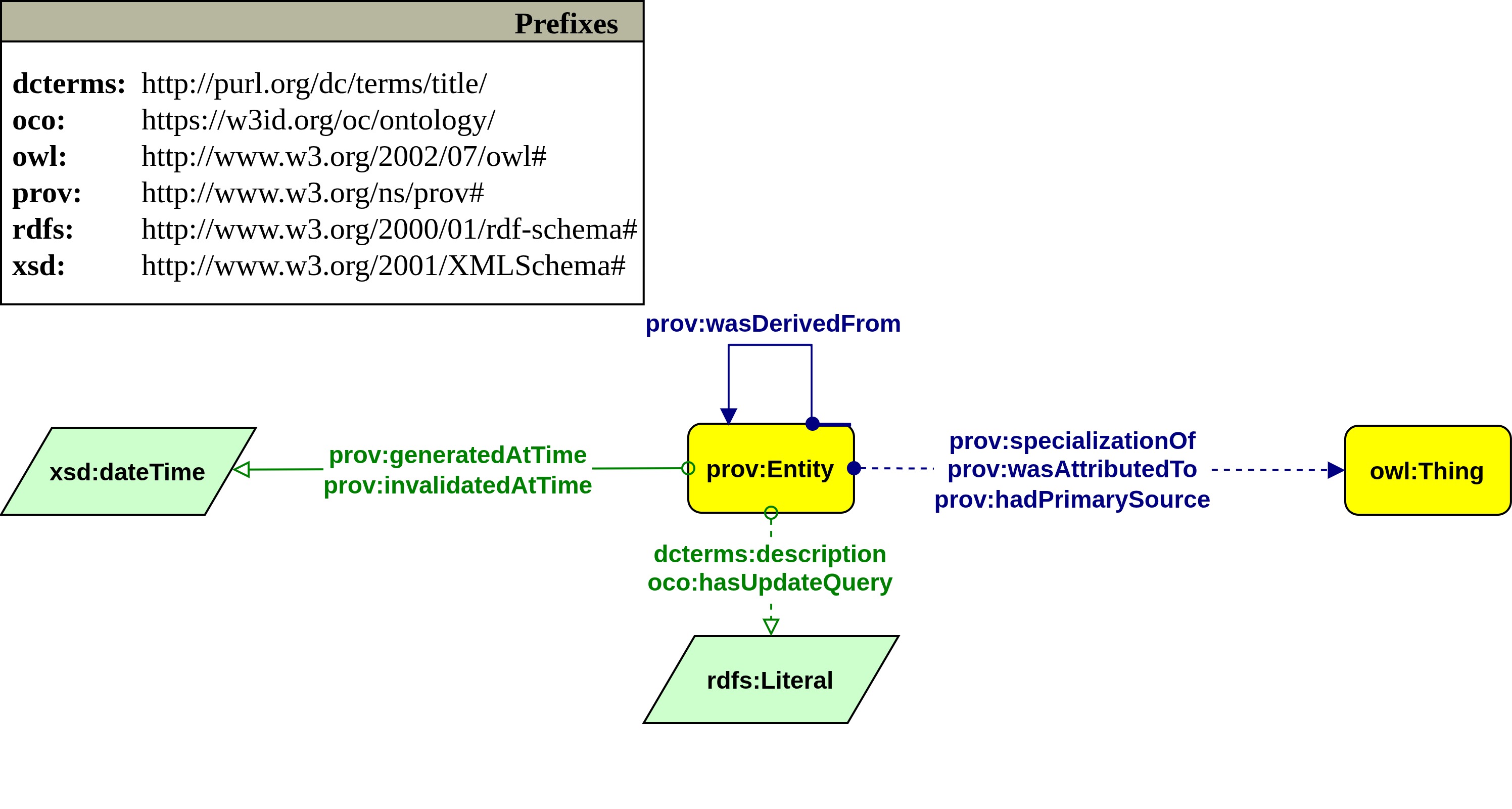}
\caption{Provenance representation in the OpenCitations Data Model, shown using the Graffoo graphical framework \citep{falcoModellingOWLOntologies2014a}. Each entity snapshot records generation time, invalidation time, responsible agent, primary source, description, update query, and link to the previous snapshot via \texttt{prov:wasDerivedFrom}. The snapshot is linked to the tracked entity via \texttt{prov:specializationOf}.}
\label{fig:ocdm_provenance}
\end{figure}

The OpenCitations Data Model (OCDM) provides mechanisms for representing provenance and change tracking in RDF datasets. Its provenance layer builds on the W3C PROV Ontology (PROV-O), extending it with a single additional property: \texttt{oco:hasUpdateQuery} \citep{peroniDocumentinspiredWayTracking2016}, which records SPARQL \texttt{INSERT DATA} and \texttt{DELETE DATA} operations between entity snapshots. Each entity's change history is stored in a provenance named graph as a chain of snapshots linked via \texttt{prov:wasDerivedFrom}. These update queries must contain only absolute URIs and literals, while prefixes and variables are not permitted. Figure~\ref{fig:ocdm_provenance} illustrates the OCDM provenance structure.

\begin{figure}[htbp]
\centering
\includegraphics[width=\textwidth]{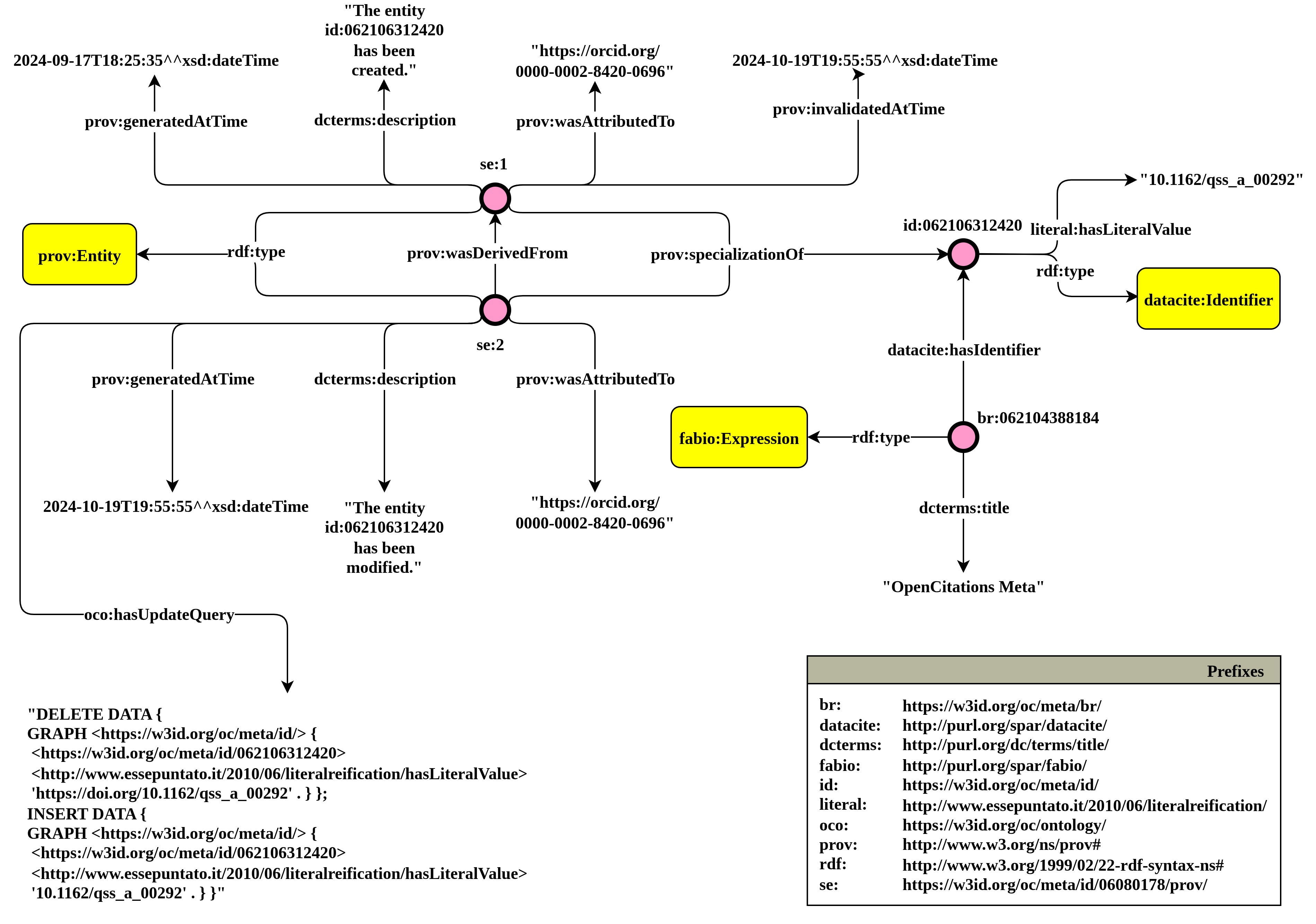}
\caption{Usage example of the OpenCitations Data Model. The identifier entity was initially registered with an incorrect DOI, which was subsequently corrected by an agent. The correction generated a new snapshot linked to the previous one via \texttt{prov:wasDerivedFrom}.}
\label{fig:ocdm_usage}
\end{figure}

\begin{sloppypar}
Figure~\ref{fig:ocdm_usage} shows a usage example of the OpenCitations Data Model. \texttt{id:062106312420} is an identifier of the bibliographic resource \texttt{br:062104388184}, whose title is ``OpenCitations Meta''. The identifier was initially registered with an incorrect DOI, that is, ``\url{https://doi.org/10.1162/qss_a_00292}'' instead of ``10.1162/qss\_a\_00292'', where the error is in the inclusion of the full URL instead of the DOI string. An agent corrected this mistake, generating a new snapshot deriving from the previous one.
\end{sloppypar}

Listing~\ref{lst:ocdm_example} shows the RDF representation in TriG syntax. The provenance named graph stores the snapshot chain. The modification snapshot (\texttt{se:2}) records the DOI correction as a \texttt{DELETE DATA}/\texttt{INSERT DATA} pair.

\begin{listing}[htbp]
\begin{lstlisting}[basicstyle=\footnotesize\ttfamily,breaklines=true,breakatwhitespace=false,showstringspaces=false]
@prefix br: <https://w3id.org/oc/meta/br/>.
@prefix datacite: <http://purl.org/spar/datacite/>.
@prefix dcterms: <http://purl.org/dc/terms/>.
@prefix fabio: <http://purl.org/spar/fabio/>.
@prefix id: <https://w3id.org/oc/meta/id/>.
@prefix literal: <http://www.essepuntato.it/2010/06/literalreification/>.
@prefix oco: <https://w3id.org/oc/ontology/>.
@prefix prov: <http://www.w3.org/ns/prov#>.
@prefix se: <https://w3id.org/oc/meta/id/062106312420/prov/se/>.
@prefix xsd: <http://www.w3.org/2001/XMLSchema#>.

<https://w3id.org/oc/meta/br/> {
    br:062104388184 a fabio:Expression;
        dcterms:title "OpenCitations Meta";
        datacite:hasIdentifier id:062106312420.
}

<https://w3id.org/oc/meta/id/> {
    id:062106312420 a datacite:Identifier;
        datacite:usesIdentifierScheme datacite:doi;
        literal:hasLiteralValue "10.1162/qss_a_00292".
}

<https://w3id.org/oc/meta/id/062106312420/prov/> {
    se:2 a prov:Entity;
        oco:hasUpdateQuery """
            DELETE DATA { GRAPH <https://w3id.org/oc/meta/id/> {
                <https://w3id.org/oc/meta/id/062106312420>
                    <http://www.essepuntato.it/2010/06/literalreification/hasLiteralValue>
                    "https://doi.org/10.1162/qss_a_00292" . } };
            INSERT DATA { GRAPH <https://w3id.org/oc/meta/id/> {
                <https://w3id.org/oc/meta/id/062106312420>
                    <http://www.essepuntato.it/2010/06/literalreification/hasLiteralValue>
                    "10.1162/qss_a_00292" . } }"""^^xsd:string;
        prov:generatedAtTime "2024-10-19T19:55:55"^^xsd:dateTime;
        prov:specializationOf id:062106312420;
        prov:wasAttributedTo <https://orcid.org/0000-0002-8420-0696>;
        prov:wasDerivedFrom se:1.

    se:1 a prov:Entity;
        prov:generatedAtTime "2024-10-10T23:44:45"^^xsd:dateTime;
        prov:hadPrimarySource <https://api.crossref.org/works/10.1162/qss_a_00292>;
        prov:invalidatedAtTime "2024-10-19T19:55:55"^^xsd:dateTime;
        prov:specializationOf id:062106312420;
        prov:wasAttributedTo <https://orcid.org/0000-0002-8420-0696>.
}
\end{lstlisting}
\caption{Usage example of the OpenCitations Data Model in TriG syntax.}
\label{lst:ocdm_example}
\end{listing}

\section{The Time Agnostic Library}
\label{sec:time-agnostic-library}

The Time Agnostic Library implements algorithms that use the OCDM provenance specification to support the temporal query types identified in the taxonomy by \citet{fernandezEvaluatingQueryStorage2016}. While this taxonomy defines temporal operations at the level of full dataset versions, the OCDM provenance model tracks changes at the entity level, where each entity maintains its own chain of provenance snapshots. The algorithms adapt the taxonomy accordingly, reconstructing individual entity states rather than complete dataset snapshots.

\subsection{Version and delta materialization}

Version materialization is the foundational operation that underpins all temporal query types. It reconstructs past entity states by inverting the SPARQL UPDATE operations stored in the OCDM provenance graph. The basic building block is Algorithm~\ref{alg:apply_inverse_update}, which takes a set of quads representing the current entity state and a single SPARQL UPDATE string, and modifies the quad set to reflect the inverse of that update. The algorithm parses the update query to extract \texttt{DELETE DATA} and \texttt{INSERT DATA} operations, then inverts each action. Quads from \texttt{DELETE DATA} blocks are added back to the current state, since they were removed in the forward direction, while quads from \texttt{INSERT DATA} blocks are discarded, since they were added in the forward direction.

\begin{algorithm}[htbp]
\KwInput{A set of quads $currentState$, a SPARQL UPDATE string $update$}
\KwOutput{$currentState$ modified in place to reflect the inverse of $update$}
$operations \leftarrow \textsc{ParseUpdate}(update)$\;
\ForEach{$(type, quads) \in operations$}{
    \If{$type = \texttt{DeleteData}$}{
        \ForEach{$quad \in quads$}{
            $currentState \leftarrow currentState \cup \{quad\}$\;
        }
    }
    \ElseIf{$type = \texttt{InsertData}$}{
        \ForEach{$quad \in quads$}{
            $currentState \leftarrow currentState \setminus \{quad\}$\;
        }
    }
}
\caption{\textsc{ApplyInverseUpdate}: invert a SPARQL UPDATE on a quad set.}
\label{alg:apply_inverse_update}
\end{algorithm}

For instance, applying Algorithm~\ref{alg:apply_inverse_update} to the update query in Listing~\ref{lst:ocdm_example} re-inserts the deleted DOI URL and removes the corrected value, restoring the entity to its state at $t_{n-1}$.

Algorithm~\ref{alg:version_materialization} builds on Algorithm~\ref{alg:apply_inverse_update} to reconstruct the full version history of an entity. Given an entity URI and an optional time interval, it produces a map from timestamps to quad sets, each representing the entity's state at that point in time. The algorithm retrieves all provenance snapshots of the entity sorted by descending time, then walks backward from the most recent snapshot through the entire chain, applying one inverse update at each step. When a time interval is provided, only the states at timestamps within the interval are retained; when no interval is specified, all timestamps are collected, yielding the complete entity history. If an interval is given but no snapshot falls within it, the algorithm falls back to the most recent snapshot before the interval starts. Because each update is applied exactly once regardless of how many target timestamps are requested, this approach avoids the redundant reapplication that would result from independently reconstructing each target from the current state.

\begin{algorithm}[htbp]
\KwInput{Entity URI $entity$, optional time interval $[t_s, t_e]$}
\KwOutput{A map from timestamps to quad sets}
$snapshots \leftarrow \textsc{SortDesc}(\textsc{QueryProvSnapshots}(entity),\; time)$\;
\lIf{$snapshots = \emptyset$}{\Return{$\emptyset$}}
\uIf{$[t_s, t_e] \neq \texttt{null}$}{
    $targetTimes \leftarrow \textsc{FilterByInterval}(snapshots, [t_s, t_e])$\;
    \If{$targetTimes = \emptyset$ \textbf{and} $t_s \neq \texttt{null}$}{
        $priorSnapshots \leftarrow \{snap \in snapshots \mid snap.time \le t_s\}$\;
        \lIf{$priorSnapshots = \emptyset$}{\Return{$\emptyset$}}
        $targetTimes \leftarrow \{\arg\max_{snap \in priorSnapshots} snap.time\}$\;
    }
}
\Else{
    $targetTimes \leftarrow \{snap.time \mid snap \in snapshots\}$\;
}
$currentState \leftarrow \textsc{QueryDataset}(entity)$\;
$result \leftarrow \emptyset$\;
\For{$i \leftarrow 0$ \KwTo $|snapshots| - 1$}{
    \If{$i > 0$ \textbf{and} $snapshots[i-1].updateQuery \neq \texttt{null}$}{
        \textsc{ApplyInverseUpdate}$(currentState, snapshots[i-1].updateQuery)$\;
    }
    \If{$snapshots[i].time \in targetTimes$}{
        $result[snapshots[i].time] \leftarrow \text{copy}(currentState)$\;
    }
}
\Return{$result$}
\caption{\textsc{VersionMaterialization}: incremental entity state reconstruction.}
\label{alg:version_materialization}
\end{algorithm}

Algorithm~\ref{alg:delta_materialization} performs delta materialization: given a list of SPARQL UPDATE strings, it produces two sets $\Delta^+$ and $\Delta^-$ representing the net additions and deletions. For each quad in a \texttt{DELETE DATA} block, if it was previously recorded as an addition, the two cancel out; otherwise, it is added to $\Delta^-$. Symmetrically, each \texttt{INSERT DATA} quad cancels a prior deletion or is added to $\Delta^+$. This composition yields the minimal net change across the interval.

\begin{algorithm}[htbp]
\KwInput{A list of SPARQL UPDATE strings $updates$}
\KwOutput{$(\Delta^+, \Delta^-)$: net additions and deletions}
$\Delta^+ \leftarrow \emptyset$\;
$\Delta^- \leftarrow \emptyset$\;
\ForEach{$update \in updates$}{
    \ForEach{$(type, quads) \in \textsc{ParseUpdate}(update)$}{
        \uIf{$type = \texttt{DeleteData}$}{
            \ForEach{$quad \in quads$}{
                \lIf{$quad \in \Delta^+$}{$\Delta^+ \leftarrow \Delta^+ \setminus \{quad\}$}
                \lElse{$\Delta^- \leftarrow \Delta^- \cup \{quad\}$}
            }
        }
        \ElseIf{$type = \texttt{InsertData}$}{
            \ForEach{$quad \in quads$}{
                \lIf{$quad \in \Delta^-$}{$\Delta^- \leftarrow \Delta^- \setminus \{quad\}$}
                \lElse{$\Delta^+ \leftarrow \Delta^+ \cup \{quad\}$}
            }
        }
    }
}
\Return{$(\Delta^+, \Delta^-)$}
\caption{\textsc{DeltaMaterialization}: compose update queries into net changes.}
\label{alg:delta_materialization}
\end{algorithm}

\subsection{Single and cross-version structured queries}

A naive approach to temporal SPARQL queries would reconstruct every version of the entire dataset and evaluate the query against each one. Since the system operates live on a production triplestore, materializing the full dataset history at query time would be prohibitively expensive. The approach adopted here reconstructs only the minimal subset of entities sufficient to answer the query. In the provenance model, each entity's change history is indexed by its subject IRI, so the algorithm must first identify which subject IRIs are relevant to the query patterns. To this end, it classifies triple patterns according to whether they provide direct or indirect access to a subject-position IRI.

\textbf{Definition 4 (Anchored and unanchored triple pattern).} A triple pattern $(s, p, o)$ is anchored if its subject is an IRI, or if any variable in the pattern can be traced to a subject-position IRI through a chain of object-to-subject variable links across the query's triple patterns.

Assume pairwise disjoint infinite sets $I$, $V$, and $L$ (IRIs, Variables, Literals). For a query $Q$, define $\text{Reachable}(v)$ for a variable $v \in V$ recursively: $\text{Reachable}(v) \Leftrightarrow \exists (s', p', o') \in Q : o' = v \land (s' \in I \lor (s' \in V \land \text{Reachable}(s')))$, with $(s, p, o) \in (I \cup V) \times (I \cup V) \times (I \cup L \cup V)$. Then Anchored$(s, p, o) \Leftrightarrow s \in I \lor \exists v \in \{s, p, o\} \cap V : \text{Reachable}(v)$.

Conversely, Unanchored$(s, p, o) \Leftrightarrow \neg$Anchored$(s, p, o)$: no variable in the pattern is reachable from a subject-position IRI through the object-to-subject chain.

Listing~\ref{lst:anchored_patterns} shows an example of an anchored pattern query. The URI \texttt{br:\allowbreak 062104388184} serves as the subject-position anchor. The algorithm reconstructs its version history, then discovers the identifier entities through the object variable \texttt{?id} and materializes their histories. The remaining variables \texttt{?scheme} and \texttt{?value} are resolved from the reconstructed identifier entities.

\begin{listing}[htbp]
\begin{lstlisting}[basicstyle=\small\ttfamily,breaklines=true,breakatwhitespace=false,showstringspaces=false]
PREFIX literal: <http://www.essepuntato.it/2010/06/literalreification/>
PREFIX datacite: <http://purl.org/spar/datacite/>
SELECT DISTINCT ?id ?scheme ?value
WHERE {
    <https://w3id.org/oc/meta/br/062104388184> datacite:hasIdentifier ?id.
    ?id datacite:usesIdentifierScheme ?scheme.
    ?id literal:hasLiteralValue ?value.
}
\end{lstlisting}
\caption{Example of a SPARQL query containing only anchored triple patterns.}
\label{lst:anchored_patterns}
\end{listing}

Consider instead a query to retrieve all identifiers and their literal values (Listing~\ref{lst:unanchored_pattern}). The unanchored pattern \texttt{?id literal:\allowbreak hasLiteralValue ?literal} lacks connection to any URI in the query.

\begin{listing}[htbp]
\begin{lstlisting}[basicstyle=\small\ttfamily,breaklines=true,breakatwhitespace=false,showstringspaces=false]
PREFIX literal: <http://www.essepuntato.it/2010/06/literalreification/>
SELECT ?id ?literal
WHERE {
    ?id literal:hasLiteralValue ?literal.
}
\end{lstlisting}
\caption{Example of a SPARQL query containing an unanchored triple pattern.}
\label{lst:unanchored_pattern}
\end{listing}

Regardless of the query shape, Algorithm~\ref{alg:version_query} takes a SPARQL query and an optional time interval, and returns a map from timestamps to lists of variable bindings. It follows three stages: discover which entities are affected, reconstruct their temporal states, and evaluate the query against each state.

\begin{algorithm}[htbp]
\KwInput{A SPARQL query $query$, optional time interval $[t_s, t_e]$}
\KwOutput{A map from timestamps to lists of variable bindings}
$patterns \leftarrow \textsc{ParsePatterns}(query)$\;
\tcp{Stage 1: entity discovery}
\ForEach{$pattern \in patterns$}{
    \uIf{$\textsc{Subject}(pattern) \in I$}{
        $entities \leftarrow entities \cup \{\textsc{Subject}(pattern)\}$\;
    }
    \uElseIf{$\neg$\textsc{IsAnchored}$(pattern)$}{
        $entities \leftarrow entities \cup \textsc{DiscoverEntities}(pattern)$\;
    }
}
\tcp{Stage 2: version reconstruction}
\ForEach{$entity \in entities$}{
    $versions[entity] \leftarrow \textsc{VersionMaterialization}(entity, [t_s, t_e])$\;
}
$temporalGraphs \leftarrow \textsc{AlignSnapshots}(versions)$\;
\If{\textsc{HasAnchored}$(patterns)$}{
    $temporalGraphs \leftarrow \textsc{ResolveVariables}(patterns, temporalGraphs)$\;
}
\tcp{Stage 3: query evaluation}
\ForEach{$(timestamp, graph) \in temporalGraphs$}{
    $result[timestamp] \leftarrow \textsc{EvaluatePatterns}(query, graph)$\;
}
\Return{$result$}
\caption{\textsc{VersionQuery}: temporal version query.}
\label{alg:version_query}
\end{algorithm}

In the first stage, the algorithm classifies each triple pattern. When the subject is an IRI, the entity is directly available. Anchored patterns with a variable subject contribute no entities at this stage: their variables are resolved later by \textsc{ResolveVariables}. For unanchored patterns, entity discovery combines two sources: a SPARQL query against the current dataset identifies entities that presently match the pattern, and a search through stored update queries reveals entities whose historical states matched but have since changed.

In the second stage, each discovered entity is reconstructed via the version materialization procedure described in Algorithm~\ref{alg:version_materialization}. When a time interval is specified, only the versions within that interval are materialized; otherwise, the complete history is reconstructed.

Since each entity has its own provenance chain, different entities produce snapshots at different timestamps. The \textsc{AlignSnapshots} procedure (Algorithm~\ref{alg:align_snapshots}) takes a map from entities to their per-timestamp quad sets and merges them into a single map from timestamps to unified quad sets. For each timestamp, it collects the quads of all entities that have a snapshot at that time. An entity that was last modified at $t_n$ and has no snapshot at a later timestamp $t_{n+1}$ did not change between the two: its quads from $t_n$ are carried forward to $t_{n+1}$.

\begin{algorithm}[htbp]
\KwInput{A map $versions$: entity $\rightarrow$ (timestamp $\rightarrow$ quad set)}
\KwOutput{A map $temporalGraphs$: timestamp $\rightarrow$ quad set}
\ForEach{$(entity, entitySnapshots) \in versions$}{
    \ForEach{$(timestamp, quads) \in entitySnapshots$}{
        $temporalGraphs[timestamp] \leftarrow temporalGraphs[timestamp] \cup quads$\;
    }
}
$sortedTimestamps \leftarrow \textsc{SortAsc}(\text{keys}(temporalGraphs))$\;
\For{$i \leftarrow 1$ \KwTo $|sortedTimestamps| - 1$}{
    $prevTimestamp \leftarrow sortedTimestamps[i-1]$\;
    $curTimestamp \leftarrow sortedTimestamps[i]$\;
    $subjects \leftarrow \{quad.subject \mid quad \in temporalGraphs[prevTimestamp]\}$\;
    \ForEach{$entity \in subjects$}{
        \If{$curTimestamp \notin \text{keys}(versions[entity])$}{
            $temporalGraphs[curTimestamp] \leftarrow temporalGraphs[curTimestamp] \cup \{quad \in temporalGraphs[prevTimestamp] \mid quad.subject = entity\}$\;
        }
    }
}
\Return{$temporalGraphs$}
\caption{\textsc{AlignSnapshots}: merge per-entity versions into temporal graphs.}
\label{alg:align_snapshots}
\end{algorithm}

When the query contains anchored patterns, the initial set of discovered entities may not be sufficient: variables in the query can resolve to URIs of entities that have not yet been reconstructed. The \textsc{ResolveVariables} procedure (Algorithm~\ref{alg:resolve_variables}) takes a set of triple patterns, the aligned temporal graphs, and the set of already reconstructed entities, and returns updated temporal graphs that include all newly discovered entities. It operates by iteratively binding variables and expanding the set of known entities. A pattern $(s, p, o)$ is \textit{solvable} when exactly one of its three elements is an unbound variable and that variable is in the object position. This restriction reflects the reconstruction model: the aligned graphs contain quads indexed by subject IRI, enabling object lookups for a given subject and predicate but not subject lookups for a given predicate and object, since the latter would require those subject entities to already be reconstructed. The procedure scans the patterns for solvable ones, looks up matching quads to collect possible values, materializes any newly discovered entities, realigns snapshots, substitutes the bound values back into the patterns, and repeats until no solvable patterns remain.

\begin{algorithm}[htbp]
\KwInput{A set of triple patterns $patterns$, aligned temporal graphs $temporalGraphs$, set of reconstructed entities $knownEntities$}
\KwOutput{Updated $temporalGraphs$ with newly discovered entities}
$changed \leftarrow \texttt{true}$\;
\While{$changed$}{
    $changed \leftarrow \texttt{false}$\;
    $solvable \leftarrow \{(s, p, o) \in patterns \mid o \in V \land s \notin V \land p \notin V\}$\;
    \ForEach{$(s, p, v) \in solvable$}{
        $values \leftarrow \emptyset$\;
        \ForEach{$(timestamp, graph) \in temporalGraphs$}{
            $values \leftarrow values \cup \{o \mid (s', p', o) \in graph \land s' = s \land p' = p\}$\;
        }
        \ForEach{$entity \in (values \cap I) \setminus knownEntities$}{
            $versions[entity] \leftarrow \textsc{VersionMaterialization}(entity)$\;
            $temporalGraphs \leftarrow \textsc{AlignSnapshots}(temporalGraphs, versions[entity])$\;
            $knownEntities \leftarrow knownEntities \cup \{entity\}$\;
            $changed \leftarrow \texttt{true}$\;
        }
        $patterns \leftarrow \textsc{Substitute}(patterns, v, values)$\;
    }
}
\Return{$temporalGraphs$}
\caption{\textsc{ResolveVariables}: iterative variable binding and entity discovery.}
\label{alg:resolve_variables}
\end{algorithm}

Consider the anchored query in Listing~\ref{lst:anchored_patterns} as an example. In the first iteration, the pattern \texttt{(<br:062104388184>,\allowbreak{} datacite:hasIdentifier,\allowbreak{} ?id)} is solvable: subject and predicate are bound, and \texttt{?id} is the only variable. The procedure looks up quads with that subject and predicate across all temporal graphs, collecting the identifier URIs as values of \texttt{?id}. Each identifier URI is a new entity: its versions are materialized and the snapshots are realigned. The variable \texttt{?id} is then substituted with each concrete value, producing new patterns such as \texttt{(<id:X>,\allowbreak{} datacite:usesIdentifierScheme,\allowbreak{} ?scheme)} and \texttt{(<id:X>,\allowbreak{} literal:hasLiteralValue,\allowbreak{} ?value)}, both solvable in the next iteration. The procedure terminates when no solvable pattern remains.

In the third stage of Algorithm~\ref{alg:version_query}, the query is evaluated against each temporal graph.

\subsection{Single and cross-delta structured queries}

Delta structured queries track entity creation, modification, and deletion events. They support both single-delta queries (changes within a specific time interval) and cross-delta queries (changes across the entire dataset history). Algorithm~\ref{alg:delta_query} takes a SPARQL query and an optional time interval, and returns a map from entity URIs to change records, where each record contains the creation and deletion timestamps (if applicable) together with the net additions ($\Delta^+$) and net deletions ($\Delta^-$). Entity discovery uses the same anchored/unanchored classification described for version queries.

After entity discovery, the algorithm retrieves the provenance metadata for each entity: the chronologically ordered sequence of snapshots, each recording a timestamp and, for all snapshots after the first, the update query that produced that state. Creation corresponds to the timestamp of the earliest snapshot, provided it falls within the requested interval. Deletion is detected when the final snapshot records an invalidation timestamp. When a time interval is specified, only snapshots within that interval contribute to the change computation.

The update queries from the relevant snapshots are composed into net changes via Algorithm~\ref{alg:delta_materialization}. The composition processes operations sequentially, canceling reversals: if a triple is inserted after a prior deletion of the same triple, the deletion is removed rather than retaining both; symmetrically, a deletion cancels a prior insertion.

\begin{algorithm}[htbp]
\KwInput{A SPARQL query $query$, optional time interval $[t_s, t_e]$}
\KwOutput{A map from entity URIs to change records}
\tcp{Phase 1: entity discovery (same as Algorithm~\ref{alg:version_query}, Stage 1)}
$entities \leftarrow$ discover entities from $query$ using anchored/unanchored classification\;
\tcp{Phase 2: change analysis}
$result \leftarrow \emptyset$\;
\ForEach{$entity \in entities$}{
    $snapshots \leftarrow \textsc{QueryProvenance}(entity)$\;
    \If{$snapshots = \emptyset$}{\textbf{continue}}
    $sortedSnapshots \leftarrow \textsc{SortAsc}(snapshots,\; time)$\;
    $relevantSnapshots \leftarrow \textsc{FilterByInterval}(sortedSnapshots, [t_s, t_e])$\;
    \If{$relevantSnapshots = \emptyset$}{\textbf{continue}}
    $record \leftarrow$ \{created: $\texttt{null}$, deleted: $\texttt{null}$, $\Delta^+$: $\emptyset$, $\Delta^-$: $\emptyset$\}\;
    \If{$sortedSnapshots[0] \in relevantSnapshots$}{
        $record.created \leftarrow sortedSnapshots[0].time$\;
    }
    \If{$sortedSnapshots[\text{last}].invalidated$}{
        $record.deleted \leftarrow sortedSnapshots[\text{last}].time$\;
    }
    $updateQueries \leftarrow [snap.updateQuery \mid snap \in relevantSnapshots,\; snap \neq sortedSnapshots[0],\; snap.updateQuery \neq \texttt{null}]$\;
    $(record.\Delta^+, record.\Delta^-) \leftarrow \textsc{ComposeUpdates}(updateQueries)$\;
    $result[entity] \leftarrow record$\;
}
\Return{$result$}
\caption{\textsc{DeltaQuery}: entity change identification.}
\label{alg:delta_query}
\end{algorithm}

\section{Implementation and evaluation}
\label{sec:evaluation}

\subsection{Implementation}

This methodology was implemented in a Python package \citep{arcangelomassariOpencitationsTimeagnosticlibrary7002026}, distributed as open-source software under the ISC license. It provides three classes: \texttt{Agnostic\-Entity} for VM and DM; \texttt{Version\-Query} for SV and CV; and \texttt{Delta\-Query} for SD and CD. Each class accepts an optional time interval as a tuple \texttt{(START,\,END)}: when specified, the operation targets a specific time point or version pair; when omitted, it spans the entire available history. In this way, each of the six retrieval needs considered in the taxonomy can be accomplished.

The package was tested on Blazegraph \citep{thompsonBigdataRDFGraph2014}, GraphDB \citep{bishopOWLIMFamilyScalable2011}, Apache Jena Fuseki \citep{carrollJenaImplementingSemantic2004}, OpenLink Virtuoso \citep{erlingRDFSupportVirtuoso2009}, and QLever \citep{bastQLeverQueryEngine2017}.
Test-Driven Development (TDD) \citep{beckTestdrivenDevelopmentExample2003} was adopted, achieving 100\% line coverage. Integration tests complement the TDD test suite by exercising the system under realistic usage scenarios, testing state combinations that unit tests may not reach. A Continuous Integration (CI) pipeline \citep{humbleContinuousDeliveryReliable2011} executes the full test suite on every commit via GitHub Actions, using a matrix of four Python versions (3.10 through 3.13) and all five supported triplestores, each deployed as a Docker container, for a total of 20 test configurations per commit.

\subsection{Benchmark dataset and setup}

The evaluation uses the BEAR-B benchmark \citep{fernandezEvaluatingQueryStorage2016}. BEAR-B was compiled from DBpedia Live changesets collected over three months (August to October 2015), capturing consecutive snapshots of the 100 most volatile resources at three temporal granularities: instant (21,046 versions), hourly (1,299 versions), and daily (89 versions). The dataset grows from 33,502 triples in version~0 to 43,907 triples in version~57. This evaluation uses two granularities: BEAR-B-daily and BEAR-B-hourly. Both share the same 62 query patterns: 49 with a known predicate (?P? patterns, such as \texttt{?s rdfs:label ?o}) and 13 with a known predicate and object (?PO patterns, such as \texttt{?s rdf:type dbo:Film}). Three query types are evaluated. The BEAR benchmark defines these operations as query atoms, formal functions that take a SPARQL query $Q$ as parameter: $Mat(Q, V_i)$ evaluates $Q$ at version $V_i$, $Diff(Q, V_i, V_j)$ computes the difference in $Q$'s results between two versions, and $Ver(Q)$ retrieves $Q$'s results annotated with the versions in which they hold. As noted in \hyperref[sec:temporal-queries]{Section~2}, these query atoms share names with the retrieval needs but denote different operations: for instance, the query atom called ``version materialization'' ($Mat$) evaluates a query on a single version, which at the retrieval needs level is a single-version structured query, not a version materialization. To maintain consistency with the rest of the article, the three BEAR operations are referred to here by the corresponding retrieval needs names:

\begin{itemize}
    \item Single-version structured query (SV), corresponding to BEAR's $Mat(Q, V_i)$: retrieve the triples matching a pattern at a specific version
    \item Single-delta structured query (SD), corresponding to BEAR's $Diff(Q, V_i, V_j)$: retrieve the triples added or removed between two versions for a given pattern
    \item Cross-version structured query (CV), corresponding to BEAR's $Ver(Q)$: retrieve the triples matching a pattern across all versions
\end{itemize}

SV queries were executed at every version for each of the 62 patterns. SD queries were executed at 12 version pairs per pattern: all pairs start from version~0, and the end versions are 11 equally spaced points through the version history (at intervals of 5 versions on BEAR-B-daily and 100 on BEAR-B-hourly), plus the final version as the 12th endpoint. CV queries were executed once per pattern. Each query was repeated 5 times.

For the Time Agnostic Library, the BEAR-B data was converted to the OCDM provenance format. Among the four formats distributed by BEAR-B (IC, CB, TB, and CBTB, corresponding to the archiving policies described in \hyperref[sec:storage-paradigms]{Section 3}), IC and CB were considered as input for the conversion. A reusable conversion module was developed and integrated in the library to produce OCDM-compliant N-Quads from either IC or CB input. CB conversion is faster than IC because it uses deltas directly rather than computing diffs between consecutive snapshots (0.4~seconds versus 6.6~seconds on BEAR-B-daily, 1.1~seconds versus 106.2~seconds on BEAR-B-hourly). However, the CB deltas distributed by BEAR are inconsistent with the IC snapshots: in our experiments, the expected query results published with the benchmark could only be reproduced when computing deltas manually as diffs between consecutive IC snapshots, while ingesting the CB deltas directly produced different results, suggesting that the distributed CB files may contain errors. The IC strategy was therefore used for the OCDM conversion. The resulting N-Quads files were indexed in QLever. The total preprocessing time, including OCDM conversion and QLever indexing, was 8.0~seconds on BEAR-B-daily (2.4~seconds with CB) and 108.2~seconds on BEAR-B-hourly (3.1~seconds with CB).

For comparison, OSTRICH \citep{taelmanOSTRICHVersionedRandomAccess2018} was executed on the same hardware using its Docker container built from the branch of the source repository that implements multi-snapshot strategies \citep{pelgrinExpressiveQueryingScalable2025}. Both datasets were ingested using the \texttt{interval~5} strategy, which creates a new HDT snapshot every 5 versions. This configuration was selected because it yields the fastest ingestion time among all strategies evaluated by \citet{pelgrinExpressiveQueryingScalable2025}, completing in 10.7~seconds on BEAR-B-daily and 89.9~seconds on BEAR-B-hourly.

R43ples \citep{graubeOpenSemanticRevision2016} was also benchmarked on BEAR-B-daily, as it is the system closest in architecture to the Time Agnostic Library among those reviewed in \hyperref[sec:storage-paradigms]{Section~3}: both store provenance metadata in RDF using PROV-O, both operate on data that resides in a live triplestore, and both reconstruct past states via query rewriting rather than relying on a precomputed index. BEAR-B-daily was ingested in 112~seconds into R43ples' underlying Jena TDB store.

All experiments were conducted on a machine with an Intel Core i9-12900K (24 cores), 128~GB DDR RAM, and SSD storage, running Arch Linux with kernel 6.18.3. Memory consumption for the Time Agnostic Library was measured using Python's \texttt{tracemalloc} module, which tracks peak heap allocation within the interpreter during each query execution. Because the Time Agnostic Library is a Python library that runs in a separate process from the triplestore, \texttt{tracemalloc} captures only the memory used for version reconstruction, without including the triplestore. OSTRICH and R43ples do not allow the same separation: both are self-contained systems in which the query engine and the storage backend run within a single process, and both were executed inside Docker containers. Any memory measurement at the container or process level would reflect the combined footprint of storage, query engine, and runtime, and could not be compared with the per-query heap allocation reported for the Time Agnostic Library.

\subsection{Results}

\begin{figure}[!htb]
\centering
\includegraphics[width=\textwidth]{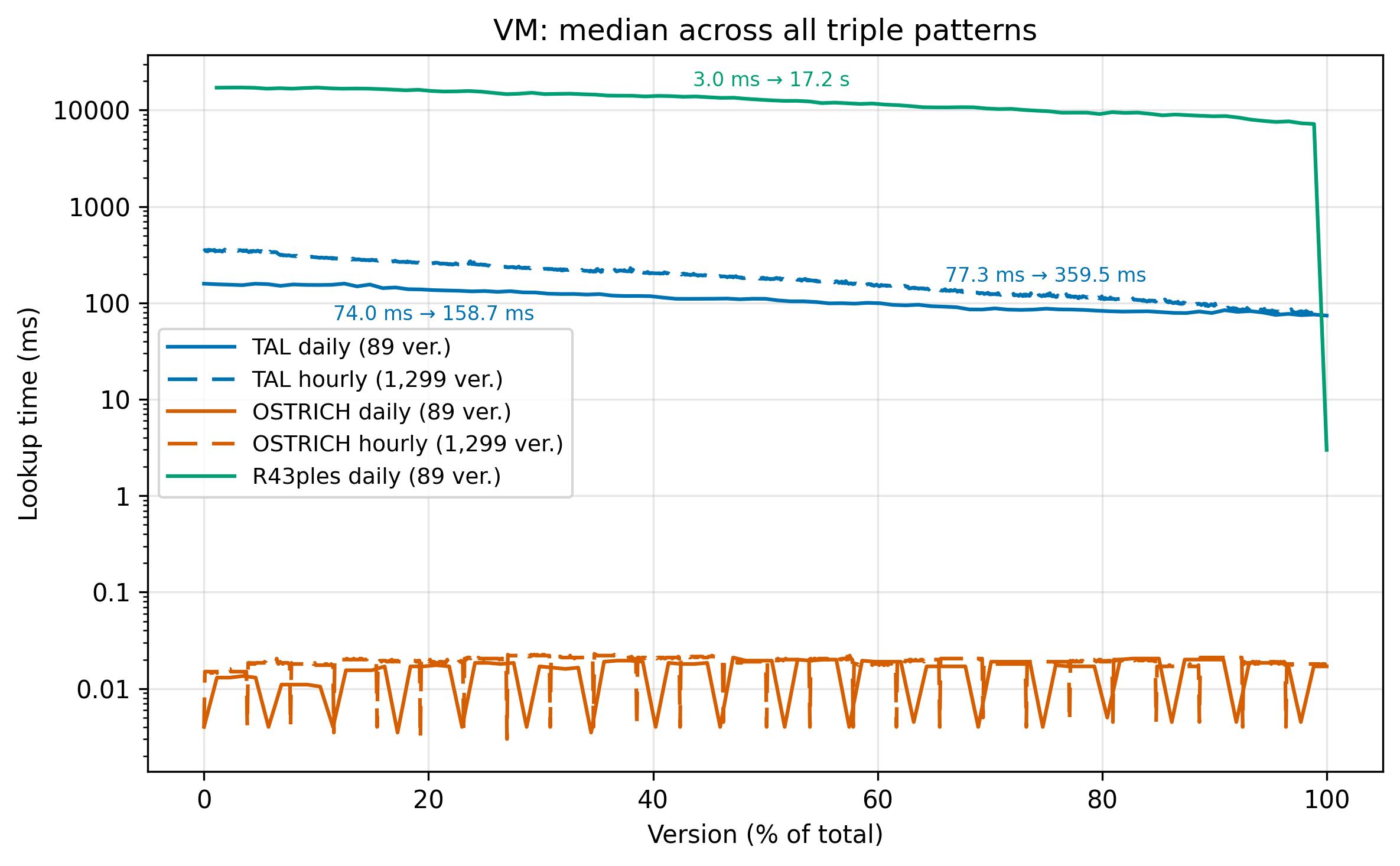}
\caption{Single-version structured query (SV): median execution time per version across all 62 triple patterns on BEAR-B-daily and BEAR-B-hourly, log scale. R43ples was benchmarked only on BEAR-B-daily.}
\label{fig:sv_comparison}
\end{figure}

\begin{figure}[!htb]
\centering
\includegraphics[width=\textwidth]{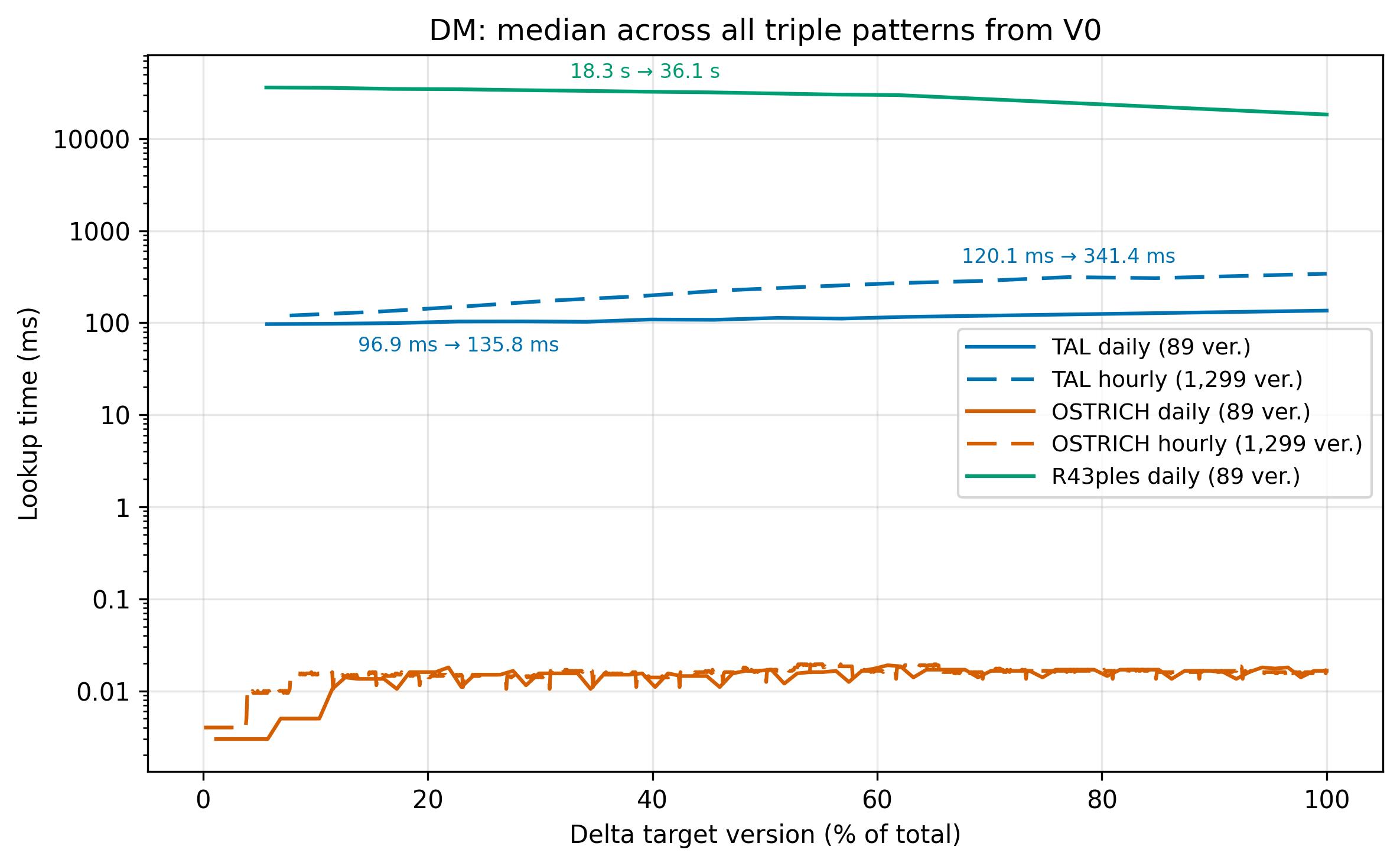}
\caption{Single-delta structured query (SD): median execution time per version pair from V0 across all 62 triple patterns on BEAR-B-daily and BEAR-B-hourly, log scale. R43ples was benchmarked only on BEAR-B-daily.}
\label{fig:sd_comparison}
\end{figure}

\begin{figure}[!htb]
\centering
\includegraphics[width=0.7\textwidth]{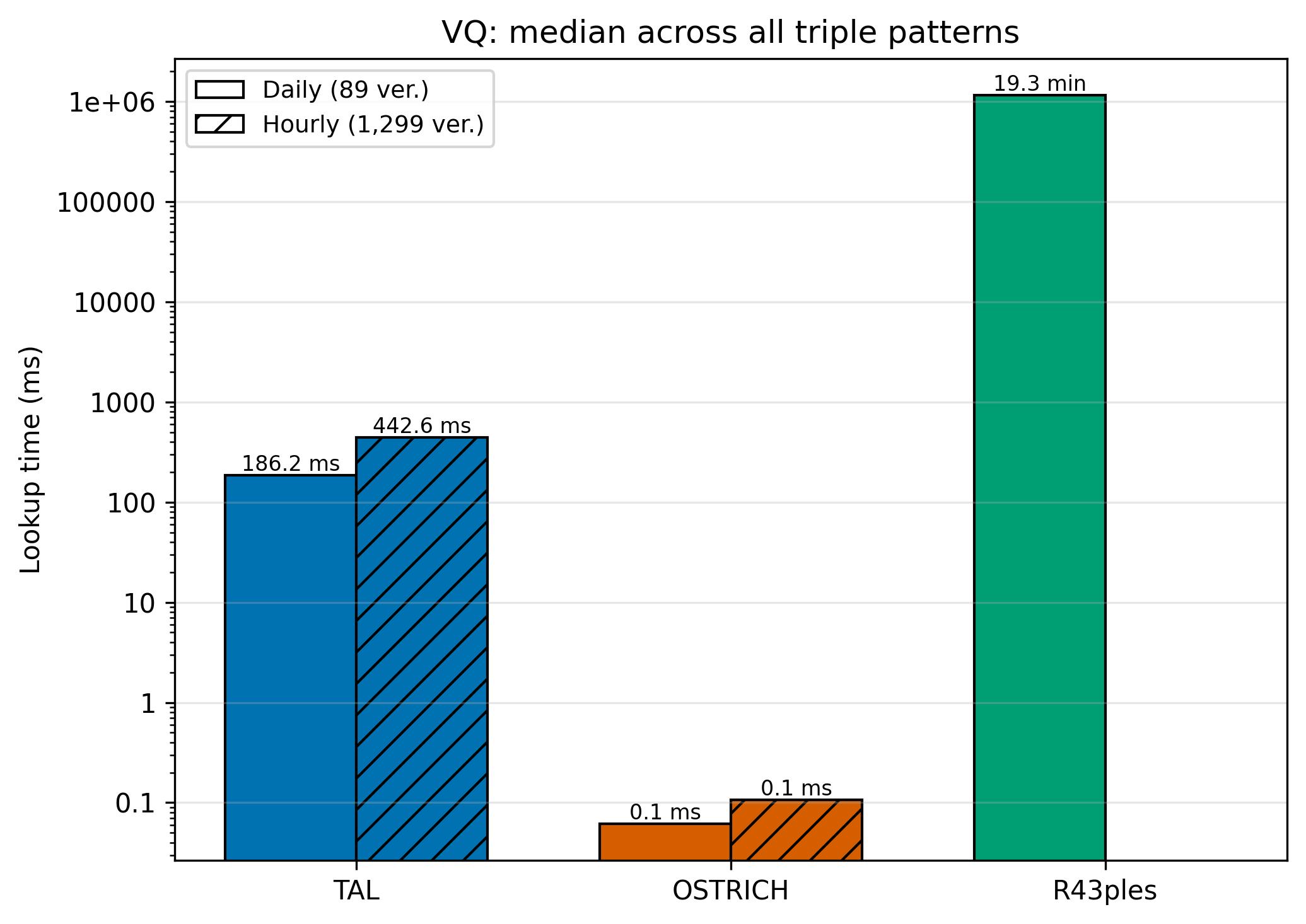}
\caption{Cross-version structured query (CV): median execution time across all 62 triple patterns on BEAR-B-daily and BEAR-B-hourly, log scale. R43ples was benchmarked only on BEAR-B-daily.}
\label{fig:cv_comparison}
\end{figure}

Tables~\ref{tab:tal_results} and~\ref{tab:comparison} summarize the execution times. Figures~\ref{fig:sv_comparison},~\ref{fig:sd_comparison}, and~\ref{fig:cv_comparison} show per-version trends across all three systems. Table~\ref{tab:memory} reports memory consumption for the Time Agnostic Library.

OSTRICH resolves all query types in sub-millisecond time on both datasets, with mean times stable across granularities (Table~\ref{tab:comparison}).

The Time Agnostic Library mean times on BEAR-B-daily are 204~ms for SV, 138~ms for SD, and 350~ms for CV (Table~\ref{tab:tal_results}). Increasing the version count from 89 to 1,299 (14.6$\times$) raises mean times by approximately 2$\times$ for SV and 3$\times$ for SD and CV, indicating sub-linear scaling with version depth. Across both datasets, ?PO patterns are approximately 3 times faster than ?P? patterns, since constraining both predicate and object reduces the number of entities to materialize.

R43ples mean times on BEAR-B-daily are 12,727~ms (12.7~seconds) for SV, 32,057~ms (32.1~seconds) for SD, and 1,174,580~ms (19.6~minutes) for CV. The SV plot (Figure~\ref{fig:sv_comparison}) shows that R43ples times decrease as the target version approaches HEAD, because R43ples walks backward from HEAD, and earlier versions require traversing more revision steps. At the latest version (HEAD), R43ples drops to a median of 3~ms, because no reconstruction is needed: the query executes directly against the current state of the underlying Jena TDB store.

Table~\ref{tab:memory} reports peak heap allocation for the Time Agnostic Library. The median allocation remains below 19~MB across both datasets and all query types. Peak values reach 396~MB for CV on BEAR-B-hourly, driven by a few patterns that reconstruct entities with large quad sets across 1,299 versions. SD has the lowest footprint (median 4~MB on BEAR-B-daily, 13~MB on BEAR-B-hourly) because it composes update queries into a net delta without materializing any version state. The ratio between hourly and daily medians (2.3$\times$ for SV, 3.0$\times$ for SD, 2.4$\times$ for CV) mirrors the scaling observed for execution times.

\begin{table}[!htb]
\centering
\caption{Time Agnostic Library execution times on BEAR-B-daily and BEAR-B-hourly.}
\label{tab:tal_results}
\begin{tabular}{llrrr}
\toprule
Dataset & Query type & Count & Mean (ms) & Median (ms) \\
\midrule
\multirow{3}{*}{Daily (89 ver.)}   & SV & 5,518  & 204 & 96 \\
                                    & SD & 744    & 138 & 107 \\
                                    & CV & 62     & 350 & 188 \\
\midrule
\multirow{3}{*}{Hourly (1,299 ver.)} & SV & 80,538 & 373 & 164 \\
                                       & SD & 744    & 404 & 201 \\
                                       & CV & 62     & 933 & 443 \\
\bottomrule
\end{tabular}
\end{table}

\begin{table}[!htb]
\centering
\caption{Performance comparison on BEAR-B, measured on the same hardware. Query times are mean values.}
\label{tab:comparison}
\begin{tabular}{lllrrrr}
\toprule
Dataset & System & Strategy & Ingestion (s) & SV (ms) & SD (ms) & CV (ms) \\
\midrule
\multirow{4}{*}{Daily}  & OSTRICH & interval 5 & 10.7  & 0.09 & 0.10 & 0.12 \\
                         & R43ples & ---        & 112.3 & 12,727 & 32,057 & 1,174,580 \\
                         & TAL     & IC         & 8.0   & 204  & 138  & 350 \\
                         & TAL     & CB         & 2.4   & \multicolumn{3}{c}{---} \\
\midrule
\multirow{3}{*}{Hourly} & OSTRICH & interval 5 & 89.9  & 0.12 & 0.11 & 0.17 \\
                         & TAL     & IC         & 108.2 & 373  & 404  & 933 \\
                         & TAL     & CB         & 3.1   & \multicolumn{3}{c}{---} \\
\bottomrule
\end{tabular}
\end{table}

\begin{table}[!htb]
\centering
\caption{Time Agnostic Library peak heap allocation on BEAR-B.}
\label{tab:memory}
\begin{tabular}{llrr}
\toprule
Dataset & Query type & Median (MB) & Max (MB) \\
\midrule
\multirow{3}{*}{Daily (89 ver.)}     & SV & 8  & 109 \\
                                      & SD & 4  & 59 \\
                                      & CV & 8  & 123 \\
\midrule
\multirow{3}{*}{Hourly (1,299 ver.)} & SV & 18 & 258 \\
                                      & SD & 13 & 184 \\
                                      & CV & 19 & 396 \\
\bottomrule
\end{tabular}
\end{table}

\subsection{Discussion and conclusion}

The performance gap between OSTRICH and TAL reflects a difference in architecture: OSTRICH preprocesses all version data into a C++ index optimized for temporal lookups, while the Time Agnostic Library reconstructs historical states at query time from provenance snapshots. The comparison with R43ples quantifies the cost of the reconstruction mechanism: both systems operate live on a triplestore and store provenance in PROV-O, but R43ples issues a SPARQL query to the triplestore at each revision step, while TAL retrieves SPARQL UPDATE operations and applies them in-memory as set operations. Since TAL performs reconstruction in memory rather than delegating it to the triplestore, one might expect a trade-off between I/O latency and memory consumption. The measurements in Table~\ref{tab:memory} show that this trade-off is contained: median heap allocation stays below 19~MB across all query types and datasets. The library keeps memory low by evaluating queries incrementally as versions are reconstructed. For single triple patterns, the working set of quads is mutated in place by applying inverse update operations, and pattern matching runs on each version before advancing to the next, so the full version history is never held in memory simultaneously. R43ples was not benchmarked on BEAR-B-hourly: the full daily benchmark (6,324 queries $\times$ 5 replications) already required approximately 155~hours, per-query cost scales linearly with the revision chain length (1,299 versus 89 versions), and SV queries grow proportionally with the number of versions (80,538 versus 5,518), yielding a projected completion time exceeding 1 year without adding findings beyond what the daily results already show.

Beyond raw performance, the two systems differ in deployment constraints. As described in \hyperref[sec:storage-paradigms]{Section~3}, OSTRICH requires offline ingestion of all version data before the store can be queried, with no interface for incremental updates after ingestion. The Time Agnostic Library operates directly on any SPARQL-compliant triplestore, eliminating the need for an ingestion phase. This makes it suitable for RDF data editors such as HERITRACE \citep{massariHERITRACEUserFriendlySemantic2025}, where each change must immediately generate queryable provenance metadata.

The latencies observed are acceptable for interactive applications: even on BEAR-B-hourly with 1,299 versions, the SV median remains 164~ms and the SD median 201~ms. The sub-linear scaling observed between the two datasets suggests that the approach remains viable as version depth grows. This scaling also justifies the exclusion of BEAR-B-instant from the evaluation: the SV workload alone would require 1,304,852 queries (21,046 versions $\times$ 62 patterns), each repeated 5 times, for a projected total exceeding 1 month, and the two granularities already tested are sufficient to characterize the scaling behavior of the library.

The queries used in this evaluation are all single triple patterns. Since the library supports full basic graph patterns, a natural next step is to benchmark it on multi-pattern queries to better characterize how join complexity affects reconstruction time and memory.

Separately, the temporal reconstruction algorithms currently cover basic graph patterns and OPTIONAL clauses. FILTER, UNION, subqueries, and aggregation remain unsupported. Broadening the SPARQL 1.1 coverage of these algorithms is a second line of future work.

\section*{Data availability statement}

The Time Agnostic Library source code, the BEAR-B conversion scripts, the benchmark execution scripts, and the raw measurements that support the findings of this study are openly available on Zenodo under the ISC licence at \url{https://doi.org/10.5281/zenodo.18957419} \citep{arcangelomassariOpencitationsTimeagnosticlibrary7002026}. The BEAR-B benchmark dataset used as input is openly distributed by its original authors at \url{https://aic.ai.wu.ac.at/qadlod/bear.html} \citep{fernandezEvaluatingQueryStorage2016}.

\section*{Acknowledgments}

This work has been partially funded by the European Union's Horizon Europe framework programme under Grant Agreements No 101095129 (GraspOS Project) and No 101188018 (GRAPHIA Project).

\section*{Generative AI disclosure}

No generative AI systems were used to design the study, generate or analyse data, produce results, or draft the scientific content of this manuscript. AI-assisted tools were used only for spell-checking and minor grammar suggestions on the English text.

\section*{Conflict of interest statement}

The authors declare no conflicts of interest.

\printbibliography

\end{document}